\documentclass[12pt]{article}
\pdfoutput=1
\usepackage[centertags]{amsmath}
\usepackage{amsfonts}
\usepackage{amssymb}
\usepackage{amsthm}
\usepackage{graphicx}
\usepackage{etoolbox}
\usepackage{youngtab}
\usepackage[noadjust]{cite}
\usepackage{filecontents}

\patchcmd{\subequations}{}%
{}{}{}

\def\a{\alpha}
\def\b{\beta}
\def\g{\gamma}

\def\d{\delta}
\def\D{\Delta}
\def\e{\epsilon}
\def\ve{\varepsilon}
\def\z{\zeta}
\def\h{\eta}
\def\th{\theta}

\def\l{\lambda}

\def\m{\mu}
\def\n{\nu}

\def\r{\rho}

\def\s{\sigma}

\def\ps{\psi}
\def\Ps{\Psi}
\def\o{\omega}

\def\tf{\tfrac}
\def\de{\partial}
\def\nb{\nabla}

\def\pss{\displaystyle{\not{\!\psi\!\,}}}
\def\Pss{\displaystyle{\not{\!\Psi\!\,}}}

\def\ths{\displaystyle{\not{\!\theta}}}
\def\nbs{\displaystyle{\not{\!\nabla\!\,}}}
\def\du{\partial_u}
\def\div{\partial_u\!\cdot\!\nabla}
\def\grad{u\!\cdot\!\nabla}
\def\spin{u\!\cdot\!\partial_u}
\def\us{\displaystyle{\not{\!u\!\,}}}
\def\dus{\displaystyle{\not{\!\partial_u\!\,}}}
\def\tr{\partial_u^2}
\def\sym{u^2}

\begin{document}
\setlength{\topmargin}{-1cm} \setlength{\oddsidemargin}{0cm}
\setlength{\evensidemargin}{0cm}

\begin{titlepage}
\begin{center}
{\Large \bf Frame- and Metric-like Higher-Spin Fermions}

\vspace{20pt}

{\large Rakibur Rahman$^{a,b}$}

\vspace{12pt}
$^a$ Department of Physics, University of Dhaka, Dhaka 1000, Bangladesh\\
\vspace{6pt}
$^b$ Max-Planck-Institut f\"ur Gravitationsphysik (Albert-Einstein-Institut)\\
     Am M\"uhlenberg 1, D-14476 Potsdam-Golm, Germany\\

\end{center}
\vspace{20pt}

\begin{abstract}
Conventional descriptions of higher-spin fermionic gauge fields appear in two varieties: the Aragone-Deser-Vasiliev frame-like formulation and the Fang-Fronsdal metric-like formulation.
We review, clarify and elaborate on some essential features of these two. For frame-like free fermions in Anti-de Sitter space, one can present a gauge-invariant Lagrangian description
such that the constraints on the field and the gauge parameters mimic their flat-space counterparts. This simplifies the explicit demonstration of the equivalence of the two formulations
at the free level. We comment on the subtleties that may arise in an interacting theory.
\end{abstract}

\end{titlepage}

\newpage
\section{Introduction}

Arbitrary-spin massless particles are expected to play a crucial role in the understanding of Quantum Gravity. Lower-spin theories may be realized as low-energy limits of
spontaneously-broken higher-spin gauge theories since lower-spin symmetries are subgroups of higher-spin ones. It is believed that the tensionless limit of string theory
is a theory of higher-spin gauge fields. The study of fermionic fields is interesting in this regard because they are required by supersymmetry.

Higher-spin gauge fields can be described in the framework of two different formulations: frame-like and metric-like. The frame-like formulation generalizes the Cartan formulation
of gravity where the gauge fields are described in terms of differential forms carrying irreducible representations of the fiber Lorentz group. This is available
in Minkowski~\cite{V-flat,hygra,AD} as well as in Anti-de Sitter (AdS)~\cite{Vasiliev:1986td,Lopatin:1987hz,V-AdS,Vasiliev:2001wa} spaces.
The metric-like formulation, on the other hand, is a generalization of the metric formulation of linearized gravity~\cite{deWit}. Originally developed by
Fronsdal~\cite{Fronsdal:1978rb,Fronsdal:1978vb} and Fang--Fronsdal~\cite{FF,Fang:1979hq}, it encodes the degrees of freedom of higher-spin particles in symmetric tensors and tensor-spinors.
In this approach, the construction of a gauge-invariant action for a higher-spin field requires that the field and the gauge parameter obey some off-shell algebraic constraints
(see~\cite{Rahman:2015pzl,Campoleoni:2017vds} for a recent review). Note that the latter requirement can be avoided by recourse to other formulations~\cite{Francia:2002aa,Bekaert:2003az,Buchbinder:2004gp,Francia:2005bu,Bekaert:2006ix,Francia:2007qt,Buchbinder:2007ak,Buchbinder:2007vq,Francia:2007ee,Campoleoni:2009gs} (see
Appendix~\ref{sec:A}).

Both these approaches are geometric, albeit in different manners, in that the frame-like formulation extends Cartan geometry whereas the metric-like formulation extends Riemannian geometry.
The latter is however a particular gauge of the former just like in the case of gravity. The construction of interacting theories for higher-spin fields, fermions in particular,
appears to be in dire need of the frame-like formulation. The metric-like formulation, in contrast, seems rather clumsy in managing the non-linearities required by gauge-theoretic consistency.
Yet it has the advantage of having a simplified field content that may make some features of the interactions more transparent. Understanding the connections
between the two 
may therefore provide valuable information~\cite{Campoleoni:2012hp,Fredenhagen:2014oua,Campoleoni:2014tfa,Boulanger:2015ova}.

In this article, we will focus exclusively on higher-spin gauge fermions. These fields appear naturally in the supersymmetric versions of Vasiliev
theory~\cite{Konstein:1989ij,Vasiliev:1990vu,Vasiliev:1992av,Sezgin:1998gg,Sezgin:1998eh,Sezgin:2002ru,Engquist:2002vr} (see~\cite{Sezgin:2012ag} for a recent review) and also
in the tensionless limit of superstring theory compactified on AdS$_5\times S^5$. The frame-like formulation of gauge fermions~\cite{V-flat,AD,hygra,V-AdS} has been discussed more
recently by various authors~\cite{Alkalaev:2001mx,Alkalaev:2006hq,Sorokin:2008tf,Z,Z2,Skvortsov:2010nh}.
The Fang-Fronsdal metric-like approach for higher-spin fermions, on the other hand, has been studied in
arbitrary dimensions in Ref.~\cite{Hallowell:2005np,Metsaev:2006zy,Metsaev:2013wza}. We will consider the free theory of a spin $s=n+\tfrac{1}{2}$ massless fermionic field
in flat and AdS spaces. Although we consider Majorana fermions for simplicity, our main results are valid almost verbatim for Dirac fermions in arbitrary spacetime dimensions.
A crucial property of frame-like fermions in flat space is their shift symmetry w.r.t.~a gauge parameter which is an irreducible tensor-spinor in the fiber space with the symmetry
property of the Young diagram $\mathbb{Y}(n-1,1)$. This symmetry makes it almost manifest that the free Lagrangian is equivalent to that of the metric-like formulation~\cite{V-flat}.
In AdS space, however, the constraints on this parameter may receive nontrivial corrections which vanish in the flat limit~\cite{Sorokin:2008tf,Z}. This is tantamount to having no such
corrections provided that some appropriate mass-like terms appear in the gauge transformation. In other words, one can have a gauge-invariant Lagrangian description for frame-like fermions
in AdS space that does not deform of the flat-space constraints on the field and the gauge parameters.

The organization of this article is as follows. In the remaining of this section we spell out our notations and conventions. A review of frame-like higher-spin massless fermions in flat space
appears in Section~\ref{sec:FFF}, where we write down the free Lagrangian~\cite{Z,Skvortsov:2010nh} and discuss its gauge symmetries along with the constraints on the field and the gauge parameters.
We also show how this theory simplifies in $D=3, 4$. Section~\ref{sec:FFA} formulates the free theory in AdS space with a trivial but convenient modification of the well-known mass-like
term~\cite{Sorokin:2008tf,Z}. By virtue of judiciously-chosen terms in the gauge transformation, we ensure that the constraints on the field and the gauge parameters mimic their flat-space counterparts.
The value of the mass parameter, determined uniquely by gauge invariance, is in complete agreement with the known results~\cite{Metsaev:2013wza,Metsaev:2003cu}.
In Section~\ref{sec:Eqv}, we demonstrate explicitly the equivalence of the frame-like Lagrangian to the metric-like one at the free level. We conclude in Section~\ref{sec:remarks} with some remarks,
especially on the subtleties
that may arise in an interacting theory. An appendix summarizes the essentials of the metric-like formulation of higher-spin gauge fermions.

\subsubsection*{Conventions \& Notations}\label{subsec:convnot}
We adopt the conventions of Ref.~\cite{Freedman:2012zz}, with mostly positive metric signature $(-+\cdots+)$.
The expression $(i_1\cdots i_n)$  denotes a totally symmetric one in all the indices $i_1,\cdots,i_n$ with no normalization factor, e.g., $(i_1i_2)=i_1i_2+i_2i_1$ etc.
The totally antisymmetric expression $[i_1\cdots i_n]$ has the same normalization.
The number of terms appearing in the (anti-)symmetrization is assumed to be the possible minimum.
A prime will denote a trace w.r.t. the background metric, e.g., $A^{\prime}=\bar{g}^{\m\n}A_{\m\n}=A^{\m}_{~\m}$.
The Levi-Civita symbol is normalized as $\ve_{01\ldots D-1}= +1$, where $D$ is the spacetime dimension.

Fiber indices and world indices will respectively be denoted with lower case Roman letters and Greek letters.
Repeated indices with the same name (appearing all as either covariant or contravariant ones)
are (anti-)symmetrized with the minimum number of terms. This results in the following rules: $a(k)a=aa(k)=(k+1)a(k+1)$,~
$a(k)a(2)=a(2)a(k)=\binom{k+2}{2}\,a(k+2)$,~ $a(k)a(k')=a(k')a(k)=\binom{k+k'}{k}\,a(k+k')$ etc,
where $a(k)$ has a unit weight by convention, and so the proportionality coefficient gives the weight of the right hand side.

The $\g$-matrices satisfy the Clifford algebra: $\{\g^a,\g^b\}=+2\h^{ab}$, and $\g^{a\,\dagger}=\h^{aa}\g^a$.
Totally antisymmetric products of $\g$-matrices, $\g^{a_1\ldots a_r}=\tf{1}{r!}\g^{[a_1}\g^{a_2}\cdots\g^{a_r]}$, have unit weight.
A ``slash'' will denote a contraction with $\gamma$-matrix, e.g., $\displaystyle{\not{\!\!A\!\,}}=\g^a A_a$.

A Majorana spinor $\chi$ obeys the reality condition: $\chi^C=\chi$. Two Majorana spinors $\chi_{1,\,2}$ follow the bilinear identity:
$\bar\chi_1\g^{a_1\ldots a_r}\chi_2=t_r\,\bar\chi_2\g^{a_1\ldots a_r}\chi_1$, where a ``bar'' denotes Majorana conjugation, and $t_r=\pm 1$,
depending on the value of $r$ and spacetime dimensionality~\cite{Freedman:2012zz}.

\section{Frame-like Fermions in Flat Space}\label{sec:FFF}

In the frame-like formulation, a fermion of spin $s=n+\tfrac{1}{2}$ is described by a vielbein-like 1-form $\Psi^{a(n-1)}$, which is
a symmetric rank-($n-1$) irreducible tensor-spinor in the fiber space:
\begin{equation}\label{fr1}
\Psi^{a(n-1)}=\Psi_\m{}^{a(n-1)}dx^\m,\qquad \g_a\Psi^{ab(n-2)}=0.
\end{equation}
The Minkowski background is described by the vielbein $\bar{e}^{\,a}=\bar{e}_\m^{\,a}dx^\m$ that satisfies $\h_{ab}\bar{e}_\m^{\,a}\bar{e}_\n^{\,b}=\h_{\m\n}$, and the spin-connection
$\bar{\o}^{ab}=\bar{\o}_\m{}^{ab}dx^\m=-\bar{\o}_\m{}^{ba}dx^\m$, which fulfill the following equations:
\begin{equation}\label{fr2}
T^a\equiv d\bar{e}^{\,a}+\bar{\o}^a{}_b\bar{e}^{\,b}=0,\qquad \r^{ab}\equiv d\bar{\o}^{ab}+\bar{\o}^a{}_c\bar{\o}^{cb}=0.
\end{equation}
In the Cartesian coordinates, in particular, the solution of Eqs.~(\ref{fr2}) is given by $\bar{e}_\m^{\,a}=\d_\m^a$ and $\bar{\o}_\m{}^{ab}=0$. We will however work
with a generic coordinate system in order to facilitate the transition to AdS space. The following quantities will be useful in the subsequent discussion:
\begin{eqnarray}
  {}^*\bar{e}_{a_1}\ldots\bar{e}_{a_p}&\equiv&\tfrac{1}{(D-p)!}\,\e_{a_1\ldots a_pa_{p+1}\ldots a_D}\bar{e}^{\,a_{p+1}}\ldots\bar{e}^{\,a_D},\label{fr2a}\\
  \h^{a_1a_2|b_1b_2}&\equiv&\tfrac{1}{2}\left(\h^{a_1b_1}\h^{a_2b_2}-\h^{a_1b_2}\h^{a_2b_1}\right).\label{fr2b}
\end{eqnarray}
The frame-like free action for a Majorana gauge fermion, in arbitrary dimensions\footnote{Majorana fermions exist in $D=3,4,8,9,10$ and $11$. In dealing with such objects it is
important to assume the anti-commuting nature of fermions already at the classical level (before quantization).}, reads~\cite{Z,Skvortsov:2010nh}:
\begin{equation}\label{fr3}
S=-\tfrac{1}{2}\int\left[\bar{\Psi}_{b_1c(n-2)}\mathcal{A}^{a_1a_2a_3,\,b_1b_2}\hat{D}\Psi_{b_2}{}^{c(n-2)}\right]{}^*\bar{e}_{a_1}\bar{e}_{a_2}\bar{e}_{a_3},
\end{equation}
where $\hat{D}$ denotes the Lorentz covariant derivative, and
\begin{equation}\label{fr4}
\mathcal{A}^{a_1a_2a_3,\,b_1b_2}\equiv\tf{1}{6n}\left(\g^{a_1a_2a_3}\h^{b_1b_2}+2(n-1)\h^{b_1b_2|[a_1a_2}\g^{a_3]}\right).
\end{equation}
The action~(\ref{fr3}) enjoys the following gauge invariance:
\begin{equation}\label{gauge-fr0}
\d \Psi^{a(n-1)}=\hat{D}\z^{a(n-1)}+\bar{e}_b\l^{b,\,a(n-1)},
\end{equation}
where the 0-form gauge parameters $\z^{a(n-1)}$ and $\l^{b,\,a(n-1)}$ are irreducible tensor-spinors of rank ($n-1$) and rank $n$ respectively with the symmetry of the Young diagrams
$\mathbb{Y}(n-1)$ and $\mathbb{Y}(n-1,1)$, i.e.,
\begin{equation}\label{YD-1}
\z^{a(n-1)}\sim\begin{aligned}
&\underbrace{\begin{tabular}{|c|c|c|c|}\hline
  $\phantom{a}$&\multicolumn{2}{|c|}{$~\cdots~$}&\phantom{a}\\\hline
\end{tabular}}_{n-1}
\end{aligned}~,
\qquad\qquad
\l^{b,\,a(n-1)}\sim\overbrace{\begin{aligned}
&\begin{tabular}{|c|c|c|c|}\hline
   $\phantom{a}$&\multicolumn{2}{|c|}{$~\cdots~$}&\phantom{a}\\\hline
\end{tabular}\\[-4pt]
&\begin{tabular}{|c|}
   $\phantom{a}$\\\hline
\end{tabular}
\end{aligned}}^{n-1}~.
\end{equation}
These irreducible tensor-spinors are subject to the following constraints:
\begin{equation}\label{identity-fr1}
\g_b\z^{ba(n-2)}=0,\qquad \g_b\l^{b,\,a(n-1)}=0,\qquad \g_c\l^{b,\,ca(n-2)}=0,\qquad \l^{a,\,a(n-1)}=0.
\end{equation}

It is obvious that the action~(\ref{fr3}) is invariant, up to a total derivative term, under the gauge transformation of the parameter $\z^{a(n-1)}$, since $\hat{D}^2=0$ in flat space.
To prove the shift symmetry w.r.t. the parameter $\l^{b,\,a(n-1)}$, let us make use of the identity: $\bar{e}^c{}^*\bar{e}_{a_1}\bar{e}_{a_2}\bar{e}_{a_3}={}^*\bar{e}_{[a_1}\bar{e}_{a_2}\d_{a_3]}^c$,
so that the variation of the action can be written as
\begin{equation}\label{gauge-fr2}
\d_\l S=-3\int\left[\bar{\Psi}_{b_1}{}^{c(n-2)}\mathcal{A}^{a_1a_2a_3,\,b_1b_2}\hat{D}\l_{a_3,\,b_2c(n-2)}\right]{}^*\bar{e}_{a_1}\bar{e}_{a_2}.
\end{equation}
Now, let us take a careful look at the identity:
\begin{eqnarray}
  6n\mathcal{A}^{a_1a_2a_3,\,b_1b_2}&=&\left(\g^{a_1a_2}\h^{b_1b_2}+2(n-1)\h^{a_1a_2|b_1b_2}\right)\g^{a_3}+(n-1)\g^{[a_1}\h^{a_2]b_1}\h^{a_3b_2}\nonumber\\
  &&~~~~~~~~~~~~~~~~~~~~~~~~-\g^{[a_1}\h^{a_2]a_3}\h^{b_1b_2}-(n-1)\g^{[a_1}\h^{a_2]b_2}\h^{a_3b_1}.\label{fr5c}
\end{eqnarray}
When plugged into the gauge variation~(\ref{gauge-fr2}), the first line on the right hand side of this identity gives vanishing contribution on account of the
$\g$-trace constraints~(\ref{identity-fr1}) on the gauge parameter $\l^{b,\,a(n-1)}$. The two terms in the second line, on the other hand, cancel each other, thanks
to the property $\l^{a,\,a(n-1)}=0$. This proves the shift symmetry since $\d_\l S=0$.

Let us count the number independent of components of the parameters $\z^{a(n-1)}$ and $\l^{b,\,a(n-1)}$. Because the frame indices are $\g$-traceless,
the number of possible values each index can take is essentially ($D-1$). Then it is easy to compute the number of components of the corresponding Young diagrams~(\ref{YD-1});
they respectively turn out to be $\binom{D+n-3}{n-1}f_D$ and $(n-1)\binom{D+n-3}{n}f_D$, where
\begin{equation}\label{fD-defined} f_D\equiv2^{D/2+((-)^D-5)/4},\end{equation}
for a Majorana fermion in $D$ dimensions. On the other hand, one needs to take into account the vanishing of the trace when one contracts two indices from different rows of
$\l^{b,\,a(n-1)}$, which removes $\binom{D+n-4}{n-2}f_D$ components. Therefore, the total numbers are given by
\begin{equation}\label{z-dof} \D_\z=\binom{D+n-3}{n-1}f_D,\qquad \D_\l=(n-1)\binom{D+n-3}{n}f_D-\binom{D+n-4}{n-2}f_D.\end{equation}
This counting will be useful later on.

\subsubsection*{\underline{Special Case: $D=3$}}

The case of $D=3$ is important in the context of hypergravity theories~\cite{hygra} (see also~\cite{Troncoso} for a recent discussion). In this case,
note that the quantity ${}^*\bar{e}_{a_1}\bar{e}_{a_2}\bar{e}_{a_3}$ reduces to the Levi-Civita tensor $\e_{a_1a_2a_3}$. Furthermore, one has at one's disposal the useful
$D$-dimensional identity:
\begin{equation}\label{fr5b}
  \mathcal{A}^{a_1a_2a_3,\,b_1b_2}=\tfrac{1}{6}\g^{a_1a_2a_3}\h^{b_1b_2}+\left(\tf{n-1}{6n}\right)\g^{a_1a_2a_3b_1b_2}-\left(\tf{n-1}{12n}\right)\left(\g^{b_1}\g^{b_2}\g^{a_1a_2a_3}+\g^{a_1a_2a_3}\g^{b_1}\g^{b_2}\right).
\end{equation}
The second term on the right hand side in the above identity is zero in $D=3$, whereas the last term gives vanishing contribution because of the $\g$-trace condition on the field.
On account of the relation: $\g^{a_1a_2a_3}\e_{a_1a_2a_3}=(3!)\mathbb{I}$, therefore, the action~(\ref{fr3}) reduces to the well-known Aragone-Deser form~\cite{hygra}:
\begin{equation}\label{fr6}
S_{D=3}=-\tfrac{1}{2}\int\bar{\Psi}_{a(n-1)}\hat{D}\Psi^{a(n-1)}.
\end{equation}
On the other hand, the gauge symmetry~(\ref{gauge-fr0})--(\ref{identity-fr1}) reduces to
\begin{equation}\label{gauge3D}
\d \Psi^{a(n-1)}=\hat{D}\z^{a(n-1)},\qquad \g_b\z^{ba(n-2)}=0.
\end{equation}
This is because in $D=3$ the shift parameter $\l^{b,\,a(n-1)}$ is trivial but $\z^{a(n-1)}$ is not,
\begin{equation}\label{bagh} \D_\l=0,\qquad \D_\z=n,\end{equation}
as one can easily see from Eq.~(\ref{z-dof}).

\subsubsection*{\underline{Special Case: $D=4$}}

In this case, the quantity ${}^*\bar{e}_{a_1}\bar{e}_{a_2}\bar{e}_{a_3}$ reduces to the 1-form $\e_{a_1a_2a_3b}\bar{e}^{\,b}$, while only the first piece on the right hand side of the identity~(\ref{fr5b})
contributes. Then the dimension-dependent identity: $\g^{a_1a_2a_3}=-i\e^{a_1a_2a_3b}\g_5\g_{b}$, reduces the action~(\ref{fr3}) to
\begin{equation}\label{fr7}
S_{D=4}=-\tfrac{i}{2}\int\bar{\Psi}_{a(n-1)}\g_5\g_b\bar{e}^{\,b}\hat{D}\Psi^{a(n-1)}.
\end{equation}
Because $\D_\z=n(n+1)\neq0,~\D_\l=(n-1)(n+2)\neq0$, both the parameters $\z^{a(n-1)}$ and $\l^{b,\,a(n-1)}$ are nontrivial, and so the gauge symmetry has the full general form of~(\ref{gauge-fr0}).
The Lagrangian~(\ref{fr7}) appeared in both Ref.~\cite{V-flat} and~\cite{AD}, but only the former reference could correctly identify the gauge symmetries.

\section{Frame-like Fermions in AdS Space}\label{sec:FFA}

The AdS background is described by the vielbein $\bar{e}^{\,a}=\bar{e}_\m^{\,a}dx^\m$ that satisfies $\h_{ab}\bar{e}_\m^a\bar{e}_\n^b=\bar{g}_{\m\n}$, and the spin-connection
$\bar{\o}^{ab}=\bar{\o}_\m{}^{ab}dx^\m=-\bar{\o}_\m{}^{ba}dx^\m$, which fulfill the following equations:
\begin{equation}\label{fr2AdS}
T^a\equiv d\bar{e}^{\,a}+\bar{\o}^a{}_b\bar{e}^{\,b}=0,\qquad \r^{ab}\equiv d\bar{\o}^{ab}+\bar{\o}^a{}_c\bar{\o}^{cb}=-\frac{1}{l^2}\bar{e}^{\,a}\bar{e}^{\,b},
\end{equation}
where $l$ is the AdS radius.
Let us write the free action for a Majorana gauge fermion in AdS space by augmenting the kinetic term, already studied in the context of flat space, by a mass term:
\begin{eqnarray}
S&=&-\tfrac{1}{2}\int\left[\bar{\Psi}_{b_1c(n-2)}\mathcal{A}^{a_1a_2a_3,\,b_1b_2}\hat{D}\Psi_{b_2}{}^{c(n-2)}\right]{}^*\bar{e}_{a_1}\bar{e}_{a_2}\bar{e}_{a_3}\nonumber\\
&&-\tfrac{1}{2}\m\int\left[\bar{\Psi}_{b_1c(n-2)}\mathcal{B}^{a_1a_2,\,b_1b_2}\Psi_{b_2}{}^{c(n-2)}\right]{}^*\bar{e}_{a_1}\bar{e}_{a_2},\label{fr3Ad}
\end{eqnarray}
where $\m$ is some parameter with the dimensions of mass, to be specified later, and
\begin{equation}\label{fr4Ad}
\mathcal{B}^{a_1a_2,\,b_1b_2}\equiv\tf{1}{2n}\left[\g^{a_1a_2}\h^{b_1b_2}+2(n-1)\h^{a_1a_2|b_1b_2}-\tf{1}{2}\left(\tf{n-1}{D+2n-4}\right)\left(\g^{b_1}\g^{b_2}\g^{a_1a_2}+\g^{a_1a_2}\g^{b_1}\g^{b_2}\right)\right].
\end{equation}
Note that our choice of $\mathcal{B}^{a_1a_2,\,b_1b_2}$ differs from that of Ref.~\cite{Sorokin:2008tf,Z} by a trivial term which vanishes upon implementing
the constraint on the field. Yet this term will be useful for our purpose.

It suffices to consider, invoking another mass parameter $\tilde{\m}$, the gauge transformation:
\begin{equation}\label{gauge-fr0Ad}
\d \Psi^{a(n-1)}=\hat{D}\z^{a(n-1)}+\tilde{\m}\bar{e}_b\left[\g^b\z^{a(n-1)}-\left(\tf{2}{D+2n-4}\right)\g^a\z^{a(n-2)b}\right]+\bar{e}_b\l^{b,\,a(n-1)},
\end{equation}
which is compatible with the $\g$-trace constraint, $\g_a\Psi^{ab(n-2)}=0$, on the field without requiring any modification of the
properties~(\ref{YD-1}) and~(\ref{identity-fr1}) of the gauge parameters. In other words, the choice of this gauge transformation~(\ref{gauge-fr0Ad}) is such that the field and the gauge
parameters mimic their flat-space properties. This point is implicit in the choice made in Ref.~\cite{Sorokin:2008tf,Z}.

To see that the shift transformation w.r.t.~the parameter $\l^{b,\,a(n-1)}$ is a symmetry of the Lagrangian~(\ref{fr3Ad}), let us first note that the invariance of the kinetic term
follows exactly the flat-space logic. Then, from the variation of the mass term, we have
\begin{equation}\label{mass-var}
\d_\l S=-2\m\int\left[\bar{\Psi}_{b_1c(n-2)}\mathcal{B}^{a_1a_2,\,b_1b_2}\l_{a_2,\,b_2}{}^{c(n-2)}\right]{}^*\bar{e}_{a_1}.
\end{equation}
On account of the identity:
\begin{eqnarray}
  2n\mathcal{B}^{a_1a_2,\,b_1b_2}&=&\h^{b_1b_2}\g^{a_1}\g^{a_2}+(n-1)\h^{a_1b_1}\h^{a_2b_2}-\tf{1}{2}\left(\tf{n-1}{D+2n-4}\right)\left(\g^{a_1a_2b_1}\g^{b_2}+\h^{b_2[a_1}\g^{a_2]}\right)\nonumber\\
  &&~~~~~~~~~~~~~~~~~~~~~~~~~~~~~~~~~~~~~~~-\h^{a_1a_2}\h^{b_1b_2}-(n-1)\h^{a_1b_2}\h^{a_2b_1},\label{fr4Ad1}
\end{eqnarray}
we then see that $\d_\l S=0$. The cancellations happen in much the same way as the identity~(\ref{fr5c}) eliminates contributions from the kinetic term.

The symmetry requirement of the Lagrangian~(\ref{fr3Ad}) w.r.t.~the $\z$-transformation in~(\ref{gauge-fr0Ad}) would relate the mass parameters $\m$ and $\tilde{\m}$ to each other
and with the inverse AdS radius. There are a priori three kinds on contributions resulting from the $\z$-transformation: 2-derivative, 1-derivative and 0-derivative ones. Not surprisingly,
by virtue of the commutator formula:
\begin{equation}\label{dsq-Ad}
\hat{D}^2\z_{a(n-1)}=-\frac{1}{l^2}\bar{e}^{\,b}\bar{e}^{\,c}\left[\h_{ab}\z_{ca(n-2)}+\tf{1}{4}\g_{bc}\z_{a(n-1)}\right],
\end{equation}
the 2-derivative piece actually reduces to a 0-derivative piece. The explicit computation makes use of the identities:
$\bar{e}^{\,b}\bar{e}^{\,c}{}^*\bar{e}_{a_1}\bar{e}_{a_2}\bar{e}_{a_3}={}^*\bar{e}_{[a_1}\d^b_{a_2}\d^c_{a_3]}$ and
$\bar{e}^{\,b}{}^*\bar{e}_{a_1}\bar{e}_{a_2}={}^*\bar{e}_{[a_1}\d^b_{a_2]}$, and leads straightforwardly to
\begin{equation}\label{AdS-last0} -\tf{(D+2n-3)(D+2n-4)}{4\,n}\tf{1}{l^2}-\tf{(D-2)(D+2n-3)}{n(D+2n-4)}\m\tilde{\m}=0,\end{equation}
in order that the even-derivative terms cancel each other. Cancellation of the 1-derivative terms,
on the other hand, requires that the following condition be met:
\begin{equation}\label{mm-rel}-(D-2)\tilde{\m}-\m=0.\end{equation}
Conditions~(\ref{AdS-last0}) and~(\ref{mm-rel}) can be combined into the relation:
\begin{equation}\label{AdS-last} \m^2 l^2=\left(n+\tf{D-4}{2}\right)^2>0,\end{equation}
which gives, up to a sign, the real mass parameter $\m$ in terms of the inverse AdS radius. The parameter $\tilde{\m}$ is then also determined from Eq.~(\ref{mm-rel}).
This uniquely fixes the Lagrangian~(\ref{fr3Ad}) as well as the gauge transformation~(\ref{gauge-fr0Ad}) while the field and gauge parameters mimic their respective
flat-space properties.

The physical significance of the mass parameter $\m$ will be made clear in the next section as we work out the gauge fixed equations of motion. To proceed,
let us forgo the language of differential forms and rewrite the action~(\ref{fr3Ad}) as:
\begin{equation}\label{fr3Ad2}
S=-\tf{1}{2}\int d^Dx\,\bar{e}\,\bar{\Psi}_{\m,\,ac(n-2)}\left(6\mathcal{A}^{\m\r\n,\,ab}\hat{D}_\r+2\m \mathcal{B}^{\m\n,\,ab}\right)\Psi_{\n,\,b}{}^{c(n-2)},
\end{equation}
where $\bar{e}\equiv\det\bar{e}_\m^{\,a}$ is the determinant of the background AdS vielbein. The resulting Lagrangian equations of motion for the frame-like fermion field
$\Ps_{\m,\,a(n-1)}$ take the form:
\begin{equation}\label{eom1}
\mathcal{R}^{\m,\,a(n-1)}\equiv\left(\tf{6}{n-1}\right)\left(\mathcal{A}^{\m\r\n,\,ab}\hat{D}_\r+\tf{1}{3}\m\,\mathcal{B}^{\m\n,\,ab}\right)\Psi_{\n,\,b}{}^{a(n-2)}=0.
\end{equation}
Here, the normalization factor keeps the equations of motion well defined also for $n=1$, as we will see.
We emphasize that the equations of motion~(\ref{eom1}) are $\g$-traceless in the fiber indices, i.e.,
\begin{equation}\label{xxx} \g_b\mathcal{R}^{\m,\,ba(n-2)}=0,\end{equation}
as they should be. Actually, the very choices of $\mathcal{A}^{\m\r\n,\,ab}$ and $\mathcal{B}^{\m\n,\,ab}$ made respectively in Eqs~(\ref{fr4}) and~(\ref{fr4Ad})
were such that the action~(\ref{fr3Ad2}) manifestly has the following form:
\begin{equation}\label{fr3Ad2.5}
S=-\tf{1}{2}\int d^Dx\,\bar{e}\,\bar{\Psi}_{\m,\,a(n-1)}\mathcal{R}^{\m,\,a(n-1)}.
\end{equation}
Clearly, the equations of motion~(\ref{eom1}) share the gauge symmetries~(\ref{gauge-fr0Ad}) of the action:
\begin{equation}\label{gauge-xxx}
\d \Psi_{\m,\,a(n-1)}=\hat{D}_\m\z_{a(n-1)}+\tilde{\m}\bar{e}_\m^{\,b}\left[\g_b\z_{a(n-1)}-\left(\tf{2}{D+2n-4}\right)\g_a\z_{a(n-2)b}\right]+\bar{e}_\m^{\,b}\l_{b,\,a(n-1)}\,.
\end{equation}
In the next section we will fix these gauge symmetries to find among other things the number of physical degrees of freedom, which should match with that of a Majorana fermion
of spin $s=n+\tf{1}{2}$\,.

\section{Equivalence of Frame- \& Metric-like Formulations}\label{sec:Eqv}

The first step to establish the equivalence of the frame- and metric-like descriptions of a gauge fermion is to find a match in the respective number of local degrees of freedom.
To count this for a frame-like fermion~\cite{V-AdS}, we rewrite the equations of motion~(\ref{eom1}) exclusively in terms of world indices:
\begin{equation}\label{eom2}
\mathcal{R}^{\m,\,\a(n-1)}\equiv\left(\g^{\m\r\n}\nabla_\r+\m\g^{\m\n}\right)\Ps_{\n,}{}^{\a(n-1)}+\tf{1}{2n}\mathcal{C}^{\m\n\b,\,\a}\Ps_{\n,\,\b}{}^{\a(n-2)}=0,
\end{equation}
where $\mathcal{C}^{\m\n\b,\,\a}$ is an operator antisymmetric in the $\m,\n,\b$ indices, given by
\begin{equation}\label{eom3}
\mathcal{C}^{\m\n\b,\,\a}\equiv\left[\g^\a,\g^{\m\r\n\b}\right]\nabla_\r-\m\left\{\g^\a,\g^{\m\n\b}\right\}-\left(\tf{2}{D+2n-4}\right)\m\g^\a\g^{\m\n\b}.
\end{equation}
Some of the dynamical modes however are not physical because of gauge invariance. In order to exclude the correct number of pure gauge modes, let us rewrite the gauge
transformations~(\ref{gauge-xxx}) as:
\begin{equation}\label{gauge-xw}
\d \Psi_{\m,\,\a(n-1)}=\nabla_\m\z_{\a(n-1)}+\tilde{\m}\left[\g_\m\z_{\a(n-1)}-\left(\tf{2}{D+2n-4}\right)\g_{\a}\z_{\a(n-2)\m}\right]+\l_{\m,\,\a(n-1)}\,.
\end{equation}
Now one can use this freedom to choose the following covariant gauge:
\begin{equation}\label{gf1}
\Pss_{\a(n-1)}\equiv\g^\m\Ps_{\m,\,\a(n-1)}=0,\qquad\Longrightarrow\qquad \Ps'_{\a(n-2)}\equiv\bar{g}^{\m\n}\Ps_{\m,\,\n\a(n-2)}=0.
\end{equation}
As a consequence, the equations of motion~(\ref{eom2}) reduce to the following form:
\begin{equation}\label{eom-gf}
\left(\displaystyle{\not{\!\nabla\!\,}}-\m\right)\Ps^{\m,}{}_{\a(n-1)}-\g^\m\nabla^\n\Ps_{\n,\,\a(n-1)}+\tf{1}{2n}\mathcal{C}^{\m\n\r,}{}_\a\,\chi_{\n,\,\r\a(n-2)}=0,
\end{equation}
where $\chi_{\m,\,\a(n-1)}$ is the irreducible part of the field $\Ps_{\m,\,\a(n-1)}$ with the symmetry of the Young diagram $\mathbb{Y}(n-1,1)$, i.e., it has exactly the same properties as
the gauge parameter $\l_{\m,\,\a(n-1)}$. Its appearance in the last term of Eq.~(\ref{eom-gf}) is easy to understand. The antisymmetry property of $\mathcal{C}^{\m\n\r,\,\a}$
removes the completely symmetric part of $\Ps_{\m,\,\a(n-1)}$, while the $\g$-trace parts are trivial  by the gauge choice~(\ref{gf1}).

The condition~(\ref{gf1}) is however not a complete gauge fixing. This can be seen by taking its gauge variation, which results in the Dirac equation for $\z_{\a(n-1)}$:
\begin{equation}\label{gf2}
\d\Pss_{\a(n-1)}\equiv\left[\displaystyle{\not{\!\nabla\!\,}}-\left(\tf{D+2n-2}{D+2n-4}\right)\m\right]\z_{\a(n-1)}=0.
\end{equation}
Not only does this allow for nontrivial solutions for $\z_{\a(n-1)}$ but it also leaves $\l_{\m,\,\a(n-1)}$ completely unaffected. Therefore, one can use to freedom of the shift parameter
$\l_{\m,\,\a(n-1)}$ to further gauge fix:
\begin{equation}\label{gf3} \chi_{\m,\,\a(n-1)}=0.\end{equation}
This finally reduces the equations of motion~(\ref{eom-gf}) to the Dirac form plus the divergence constraint:
\begin{equation}\label{eom7}
  \left(\displaystyle{\not{\!\nabla\!\,}}-\m\right)\Ps_{\m,\,\a(n-1)}=0,\qquad \nabla^\m\Ps_{\m,\,\a(n-1)}=0.
\end{equation}
To exhaust the residual freedom of $\z_{\a(n-1)}$ let us choose the gauge:
\begin{equation}\label{gf9}\Ps_{0,\,\a(n-1)}=0.\end{equation}
Its is easy to see that no residual freedom of $\z_{\a(n-1)}$ is left. A would-be residual parameter must obey some screened Poisson equation with no source term,
which has no nontrivial solutions.

The count of local physical degrees of freedom is now immediate. The system~(\ref{eom7}) describes $(D-1)\D_\z$ many dynamical variables, where $\D_\z$ is given in Eq.~(\ref{z-dof}).
But the gauge choices~(\ref{gf1}), (\ref{gf3}) and~(\ref{gf9}) respectively remove $\D_\z$, $\D_\l$ and $\D_\z$ degrees of freedom. Therefore, the total number of physical degrees
of freedom is $(D-3)\D_\z-\D_\l$, which is the same as
\begin{equation}\label{dof-adv} \D_{\text{Frame}}=\binom{D+n-4}{n}f_D\,.\end{equation}
This confirms, in view of Eq.~(\ref{dofF}), that the count matches in the two formulations: $\D_\text{Frame}=\D_{\text{Metric}}$\,.

The physical significance of the mass parameter $\m$ is now clear from the Dirac equation in~(\ref{eom7}). While Eq.~(\ref{AdS-last}) says that $\m$ must be real, one may choose $\m>0$
without any loss of generality. Then,
\begin{equation}\label{AdS-last1} \m=\frac{1}{l}\left(n+\tf{D-4}{2}\right)>0.\end{equation}
Our $\m$ corresponds to the lowest value of the mass parameter $m$ for a fermion carrying a unitary irreducible representation of the AdS isometry algebra:
\begin{equation}\label{kutta}  \left(\displaystyle{\not{\!\nabla\!\,}}-m\right)\Ps_{\m,\,\a(n-1)}=0,\qquad m\geq\m>0.\end{equation}
The bound saturates for the massless representation~\cite{Metsaev:2006zy,Metsaev:2013wza,Metsaev:2003cu}, as we see.

Next we will show that the two formulations are equivalent at the level of the free Lagrangian. With this end in view, let us decompose the fermion field
$\Ps_{\m,\,\a(n-1)}$ into totally symmetric, $\g$-traceless mixed-symmetric and $\g$-trace parts:
\begin{equation}\label{decomp} \Ps_{\m,\,\a(n-1)}=\ps_{\m\a(n-1)}+\chi_{\m,\,\a(n-1)}+\g_{[\m}\th_{\a]\a(n-2)},\end{equation}
where the fields appearing on the right hand side have the symmetry of the following Young diagrams:
\begin{equation}\label{bilai1}
\ps_{\a(n)}\sim\begin{aligned}
&\underbrace{\begin{tabular}{|c|c|c|c|}\hline
  $\phantom{a}$&\multicolumn{2}{|c|}{$~\cdots~$}&\phantom{a}\\\hline
\end{tabular}}_n\,,
\end{aligned}
\quad
\chi_{\m,\,\a(n-1)}\sim\overbrace{\begin{aligned}
&\begin{tabular}{|c|c|c|c|}\hline
   $\phantom{a}$&\multicolumn{2}{|c|}{$~\cdots~$}&\phantom{a}\\\hline
\end{tabular}\\[-4pt]
&\begin{tabular}{|c|}
   $\phantom{a}$\\\hline
\end{tabular}
\end{aligned}}^{n-1}\,,
\quad
\th_{\a(n-1)}\sim\begin{aligned}
&\underbrace{\begin{tabular}{|c|c|c|c|c|}\hline
  $\phantom{a}$&\multicolumn{2}{|c|}{$~\cdots~$}&\phantom{a}\\\hline
\end{tabular}}_{n-1}\,.
\end{aligned}
\end{equation}
We have imposed irreducibility conditions on $\chi_{\m,\,\a(n-1)}$, so that it is subject to the following constraints:
\begin{equation}\label{bilai2}
\g^\m\chi_{\m,\,\a(n-1)}=0,\qquad \g^\b\chi_{\m,\,\a(n-2)\b}=0,\qquad \chi_{\a,\,\a(n-1)}=0.
\end{equation}
Of course there will be additional constraints on the fields  $\ps_{\a(n)}$ and $\th_{\a(n-1)}$ coming from the $\g$-trace condition on the
parent field $\Ps_{\m,\,\a(n-1)}$ in the $\a$-indices. To find them, let us first take a $\g$-trace of Eq.~(\ref{decomp}) in an $\alpha$-index.
This results in
\begin{equation}\label{bilai3} \pss_{\m\a(n-2)}-(D-2)\th_{\m\a(n-2)}-(n-1)\g_\m\ths_{\a(n-2)}+\g_\a\ths_{\m\a(n-3)}=0.\end{equation}
Another $\g$-trace w.r.t. the $\m$-index gives
\begin{equation}\label{bilai4}
\ps'_{\a(n-2)}-(Dn-2n+2)\ths_{\a(n-2)}-\g_\a\th'_{\a(n-3)}=0.
\end{equation}
Now a third $\g$-trace in an $\alpha$-index yields:
\begin{equation}\label{bilai5} \pss^{\,\prime}_{\a(n-3)}-(Dn+D-4)\th^{\prime}_{\a(n-3)}+\g_\a\ths^{\,\prime}_{\a(n-4)}=0.\end{equation}
On the other hand, one could also have obtained a triple $\g$-trace by first contracting the $\m$ index with an $\a$ index in Eq.~(\ref{decomp}) and then taking a $\g$ trace.
This however produces a different result:
\begin{equation}\label{bilai6} \pss^{\,\prime}_{\a(n-3)}-(D+n-4)\th^{\prime}_{\a(n-3)}+\g_\a\ths^{\,\prime}_{\a(n-4)}=0.\end{equation}
Eqs.~(\ref{bilai5}) and~(\ref{bilai6}) impose the following constraints:
\begin{equation}\label{bilai7} \pss^{\,\prime}_{\a(n-3)}=0,\qquad \th^{\prime}_{\a(n-3)}=0,\end{equation}
i.e., the symmetric rank-$n$ field $\ps_{\a(n)}$ must be triply $\g$-traceless, whereas the symmetric rank-$(n-1)$ field $\th_{\a(n-1)}$ must be traceless. This in turn results,
from Eqs.~(\ref{bilai3}) and~(\ref{bilai4}), in the following relation:
\begin{equation}\label{bilai8}
\th_{\a(n-1)}=\left(\tf{1}{D-2}\right)\left[\pss_{\a(n-1)}-\left(\tf{1}{nD-2n+2}\right)\g_\a\ps'_{\a(n-2)}\right].
\end{equation}
Finally, plugging the above expression into the decomposition~(\ref{decomp}), we obtain:
\begin{eqnarray}
\Ps_{\m,\,\a(n-1)}&=&\ps_{\m\a(n-1)}+\chi_{\m,\,\a(n-1)}+\left(\tf{1}{D-2}\right)\left[\g_{[\m}\pss_{\a]\a(n-2)}-\left(\tf{2}{Dn-2n+2}\right)\g_{\m\a}\ps'_{\a(n-2)}\right]\nonumber\\
&&~~~~~~~~~~~+\tf{1}{(D-2)(Dn-2n+2)}\left[(n-2)\g_\a\g_\m\ps'_{\a(n-2)}-2\bar{g}_{\a(2)}\ps'_{\m\a(n-3)}\right].\label{bilai9}
\end{eqnarray}
This decomposition generalizes that of Ref.~\cite{V-flat} to arbitrary dimensions.

It will be convenient to write the covariant equations of motion~(\ref{eom2}) in the following form:
\begin{equation}\label{ge1} \mathcal{R}^{\m,\,\a(n-1)}\equiv\mathcal{O}^{\m\n,\,\a(n-1)\b(n-1)}\Ps_{\n,\,\b(n-1)}=0,\end{equation}
where we have defined the operator $\mathcal O$ as:
\begin{equation}\label{ge2}
\mathcal{O}^{\m\n,\,\a(n-1)\b(n-1)}\equiv\left(\g^{\m\r\n}\nabla_\r+\m\g^{\m\n}\right)\bar{g}^{\,\a(n-1),\,\b(n-1)}+\tf{1}{2n(n-1)}\mathcal{C}^{\m\n\b,\,\a}\bar{g}^{\,\a(n-2),\,\b(n-2)},
\end{equation}
with $\bar{g}^{\,\a(k),\,\b(k)}\equiv\tf{1}{k^2}\bar{g}^{\,\a\b}\bar{g}^{\,\a\b}\ldots\bar{g}^{\,\a\b}$ (multiplicity $k$) denoting the unit-strength symmetric tensor product of $k$ background
metric tensors. This enables us to present the corresponding Lagrangian as:
\begin{equation}\label{ge3} \tf{1}{\sqrt{-\bar{g}}}\,\mathcal{L}=-\tf{1}{2}\bar{\Ps}_{\m,\,\a(n-1)}\mathcal{O}^{\m\n,\,\a(n-1)\b(n-1)}\Ps_{\n,\,\b(n-1)}\,.\end{equation}

When the decomposition~(\ref{bilai9}) is plugged into the above Lagrangian, the irreducible mixed-symmetric part $\chi_{\m,\,\a(n-1)}$ completely drops out, thanks to the shift symmetry.
The fact that the parameter $\l_{\m,\,\a(n-1)}$ enjoys exactly the same properties as $\chi_{\m,\,\a(n-1)}$ plays a crucial role in this regard. The resulting Lagrangian contains only
the completely symmetric part $\ps_{\a(n)}$ and can be viewed as a gauge-fixed version of the original Lagrangian~(\ref{ge3}) with the gauge fixing: $\chi_{\m,\,\a(n-1)}=0$. The explicit
derivation of this Lagrangian is tedious but straightforward. The calculations can however be simplified by noting that, on account of the $\g$-tracelessness of the equations of
motion~(\ref{ge1}) in the $\a$-indices, the Lagrangian splits into the sum of two pieces:
\begin{equation}\label{ge4} \tf{1}{\sqrt{-\bar{g}}}\,\mathcal{L}=-\tf{1}{2}\bar{\Xi}_{\m,\,\a(n-1)}\mathcal{O}^{\m\n,\,\a(n-1)\b(n-1)}\Xi_{\n,\,\b(n-1)}
+\tf{1}{2}\bar{\xi}_{\m,\,\a(n-2)}\g_\a\mathcal{O}^{\m\n,\,\a(n-1)\b(n-1)}\g_\b\xi_{\n,\,\b(n-2)}\,,\end{equation}
where the tensor-spinors $\Xi_{\m,\,\a(n-1)}$ and $\xi_{\m,\,\a(n-2)}$ are given by:
\begin{eqnarray}
  \Xi_{\m,\,\a(n-1)} &=& \ps_{\m\a(n-1)}+\left(\tf{1}{D-2}\right)\left[(n-1)\g_\m\pss_{\a(n-1)}-\left(\tf{2}{Dn-2n+2}\right)\bar{g}_{\m\a}\ps'_{\a(n-2)}\right],\nonumber\\
  \xi_{\m,\,\a(n-2)} &=& \left(\tf{1}{D-2}\right)\left[-\pss_{\m\a(n-2)}+\left(\tf{1}{Dn-2n+2}\right)\left(n\g_\m\ps'_{\a(n-2)}-\g_\a\ps'_{\m\a(n-3)}\right)\right].\label{ge5}
\end{eqnarray}
One can explicitly carry out the calculations to get to the following result:
\begin{eqnarray}
-\tf{2}{\sqrt{-\bar{g}}}\,\mathcal{L}&=&\bar{\ps}_{\a(n)}\left(\not{\!\nabla\!}-\m\right)\ps^{\a(n)}
+n\bar{\pss}_{\a(n-1)}\left(\not{\!\nabla\!}+\m\right)\pss^{\a(n-1)}
-2n\bar{\pss}_{\a(n-1)}\!\nabla_\m\ps^{\m\a(n-1)}\nonumber\\
&&-\tfrac{1}{4}n(n-1)\bar{\ps}'_{\a(n-2)}\left(\not{\!\nabla\!}-\m\right)\ps'^{\,\a(n-2)}
-n(n-1)\bar{\ps}'_{\a(n-2)}\nabla_\m\pss^{\,\m\a(n-2)}.\label{ge6}
\end{eqnarray}
This indeed coincides with the Lagrangian~(\ref{f00}) for a metric-like gauge fermion in AdS space. Because only the symmetric part of the parent field $\Ps_{\m,\,\a(n-1)}$
appears in this Lagrangian, the corresponding gauge symmetry is obtained simply by a total symmetrization of the indices in Eq.~(\ref{gauge-xw}). The result is:
\begin{equation}\label{gauge-xwm}
\d \ps_{\a(n)}=\tfrac{1}{n}\left(\nb_\a\z_{\a(n-1)}-\tfrac{1}{2l}\g_{\a}\z_{\a(n-1)}\right),
\end{equation}
which also matches perfectly with the metric-like gauge symmetry~(\ref{tamm2}).

This hardly comes as a surprise. The symmetric part of $\Ps_{\m,\,\a(n-1)}$ has all the characteristics of a metric-like gauge fermion; in particular it is
triple $\g$-traceless as we have shown in Eq.~(\ref{bilai7}). Moreover, it transforms w.r.t.~a symmetric $\g$-traceless gauge parameter $\z_\a(n-1)$.
The gauge-invariant Lagrangian description for such a system is unique~\cite{Hallowell:2005np,Metsaev:2006zy,Metsaev:2013wza}.
So, $\ps_{\a(n)}$ is a metric-like gauge fermion in every sense.

\section{Remarks}\label{sec:remarks}

In this article, we have elaborated on some key features of higher-spin gauge fermions and the connections between their frame- and metric-like formulations
at the free level. A gauge-invariant frame-like Lagrangian description in AdS space, with the constraints on the
fields and the gauge parameters resembling their flat-space cousins, facilitates the explicit derivation of the corresponding metric-like Lagrangian as a gauge
fixing. This derivation generalizes that of Ref.~\cite{V-flat} to AdS space and arbitrary dimensions. Although the equivalence of the frame-
and metric-like formulations at the free level may not come as a surprise, our work fills a gap in the literature.

As is well-known, the frame-like formulation packages the non-linearities in an interacting theory in a very efficient way. For higher-spin fermions this can be seen
in a very simple setup: the Aragone-Deser hypergravity~\cite{hygra}$-$a consistent gauge theory of a spin $s=n+\tf{1}{2}$ massless Majorana fermion coupled to Einstein
gravity in 3D flat space. While only fermion bilinears appear in the frame-like formulation~\cite{hygra}, the metric-like formulation will also include four-fermion couplings
that originate from integrating out the spin-connection, just like in supergravity~\cite{Freedman:2012zz}. Moreover, the fermion-bilinear terms will look more complicated
in the metric-like variables. To see this, note that with frame-like fermions the cubic cross-coupling in the covariant language has the simple form~\cite{ours}:
\begin{equation}\label{last0}
\mathcal{L}_3\sim\bar\Ps_{\m,\,\a(n-1)}\g^{\m\n\r}\g^{\s\l}\Ps_{\n,}{}^{\a(n-1)}\de_\s h_{\r\l},
\end{equation}
where $h_{\m\n}$ is the metric perturbation. Because the irreducible hook part $\chi_{\m,\,\a(n-1)}$ of the frame-like fermion is trivial in $D=3$, the
decomposition~(\ref{bilai9}) amounts to a complicated field redefinition:
\begin{equation}\label{last1}
\Ps_{\m,\,\a(n-1)}=\ps_{\m\a(n-1)}+\g_{[\m}\pss_{\a]\a(n-2)}+\left(\tf{1}{n+2}\right)\left[n\g_\a\g_\m\ps'_{\a(n-2)}-2\h_{\m\a}\ps'_{\a(n-2)}+2\h_{\a(2)}\ps'_{\m\a(n-3)}\right],
\end{equation}
where $\ps_{\a(n)}$ is the metric-like fermion. After this redefinition is performed, the cubic coupling~(\ref{last0}) will look cumbersome in terms of the metric-like fermion.
Within the metric-like formulation, it would be more difficult to construct or to prove the consistency of this cubic coupling, say using the techniques of
Ref.~\cite{Henneaux:2012wg,Henneaux:2013gba}. The fermion-bilinear cross-couplings do not stop at any finite order in the graviton fluctuations and the situation gets only worse
at higher orders, while the frame-like formulation captures all the non-linearities in a very neat way~\cite{hygra}.

In higher dimensions the difference between the two formulations becomes more drastic. The hook part of the frame-like fermion never shows up in the interacting Lagrangian because
of the deformed shift symmetry. However, there appear the so-called ``extra'' fields: a set of additional fields that arises when one tries to construct a complete set of gauge-invariant
objects (curvatures)\footnote{The extra fields are generalizations of the spin-connection. The number of extra fields depends on the spin; the higher the spin, the more are the
extra fields needed for constructing curvatures. The extra fields however do not enter the free action, and so they are not expressed in terms of physical fields via equations of motion.
}~\cite{Fradkin:1986ka}.
To understand the role of these extra fields that are absent in the free Lagrangian, one may express them in terms of the physical fields by means of appropriate constraints
implemented via Lagrange multipliers~\cite{Vasiliev:1986td,V-AdS,Fradkin:1986ka,Fradkin:1987ks,Fradkin:1986qy}. Then, up to pure gauge parts, the extra fields are given by
derivatives of the physical fields. The extra fields therefore induce higher-derivative terms in the interactions, while their absence in the free Lagrangian merely reflects
the absence of higher-derivative kinetic terms.
Explicit solution of the aforementioned constraints are difficult, and actually not needed. The main idea of the so-called Fradkin-Vasiliev
formalism~\cite{Fradkin:1986ka,Fradkin:1987ks,Fradkin:1986qy} is that one can treat the extra fields as independent variables since most of the gauge-invariant curvatures
vanish on shell.

\subsection*{Acknowledgments}

The author is grateful to N.~Boulanger, A.~Campoleoni, G.~Lucena G\'omez, M.~Henneaux, and especially to  E.~D.~Skvortsov for valuable inputs and useful comments.
He would like to thank the organizers of the 4th Mons Workshop on Higher Spin Gauge Theories (2017), during which this study was initiated.

\begin{appendix}
\numberwithin{equation}{section}
\section{Metric-like Formulation}\label{sec:A}

The metric-like formulation of gauge fermions originated in the work of Fang and Fronsdal~\cite{FF,Fang:1979hq}, who studied the massless limit of the Lagrangian for massive
higher-spin fermions. The Fang-Fronsdal Lagrangian can be derived uniquely by considering gauge invariance and supersymmetry transformations for a massless system involving
the pair of spins $\left(s, s+\tfrac{1}{2}\right)$~\cite{Curtright:1979uz}. The construction was later generalized for maximally symmetric spaces with arbitrary dimension in Ref.~\cite{Hallowell:2005np,Metsaev:2006zy,Metsaev:2013wza}. In the metric-like formulation, a spin $s=n+\tfrac{1}{2}$ gauge fermion is described by a completely symmetric rank-$n$
tensor-spinor $\ps_{\m(n)}$ in the world indices. It satisfies the triple $\g$-trace condition:
\begin{equation}\label{tg1}\pss'_{\m(n-3)}=0.\end{equation}
It is convenient to describe metric-like theories in the operator formalism, where contraction and symmetrization of indices are realized through auxiliary variables
and tensor operations are simplified in terms of operator calculus. Symmetric tensor-spinor fields are represented by:
\begin{equation}\label{field}\ps(x,u)=\tf{1}{n!}\,\ps_{\m_1\ldots\m_n}(x)\,\bar{e}^{\,\m_1}_{a_1}(x)u^{a_1}\,\ldots\,\bar{e}^{\,\m_n}_{a_n}(x)u^{a_n},\end{equation}
where $\bar{e}^{\,\m}_a(x)$ is the background vielbein and $u^a$ is an auxiliary tangent variable. The action of the covariant derivative is defined as
a differential operation involving both $x$ and $u$:
\begin{equation}\label{covD}\nabla_\m=\bar{\nabla}_\m+\bar{\o}_\m{}^{ab}u_a\tf{\de}{\de u^b},\end{equation}
where $\bar{\nabla}_\m$ is the standard covariant derivative acting on naked tensorial indices, and $\bar{\o}_\m{}^{ab}$ the background spin connection.
In what follows we work only with the contracted auxiliary variable and the associated derivative:
\begin{equation}\label{u-du} u^\m\equiv \bar{e}^{\,\m}_{a}(x)u^{a},\quad \de_u^\m\equiv \bar{e}^{\,\m a}(x)\tf{\de}{\de u^a}.\end{equation}
The vielbein postulate then implies that $[\nb_\m,u^\n]=0$~as well as $[\nb_\m,\du^\n]=0$. The commutator of covariant derivatives on a spinor function of $u$ and $\du$
will be given by:
\begin{equation}\label{commutator}[\nb_\m,\nb_\n]=R_{\m\n\r\s}(x)u^\r\du^\s+\tf{1}{4}R_{\m\n\r\s}(x)\g^{\r\s}.\end{equation}
One would have to use the following set of operators~\cite{Hallowell:2005np,Metsaev:2006zy,Metsaev:2013wza}:
\begin{equation}\label{sett} \mathbb{G}=\left\{\nbs,\,\div,\,\grad,\,\dus,\,\us,\,\tr,\,\sym,\,\spin\right\}.\end{equation}
The set comprises eight operators: the Dirac operator $\nbs$, divergence $\div$, symmetrized-gradient $\grad$, $\g$-trace $\dus$, symmetrized-$\g$ $\us$, trace $\tr$,
symmetrized-metric $\sym$ and rank $\spin$.
These operators have nontrivial commutation relations because of $[\du^\m,u^\n]=\bar{g}^{\,\m\n}$ and the non-commutativity~(\ref{commutator}) of the
covariant derivatives if the background is non-flat.

Then, the Lagrangian for a massless fermionic field in AdS space can be written as (for a Majorana fermion, certain terms in the Lagrangian are equivalent up to total
derivatives)~\cite{Metsaev:2013wza}:
\begin{eqnarray}
\tf{1}{\sqrt{-\bar{g}}}\,\mathcal{L}&=&-\tfrac{1}{2}\bar{\ps}(\ast_n)\left(\nbs-\grad\dus-\us\,\div+\us\,\nbs\dus+\tf{1}{2}\us\,\grad\,\tr+\tf{1}{2}\sym\,\div\,\dus\right)
\ps\nonumber\\&&-\tfrac{1}{2}\bar{\ps}(\ast_n)\left(-\tf{1}{4}\sym\nbs\,\tr\right)\ps+\tf{1}{2}\m\,\bar{\ps}(\ast_n)\left(1-\us\,\dus-\tf{1}{4}\sym\,\tr\right)\ps,\label{f00}
\end{eqnarray}
where the operation: $(\ast_k)\equiv\left(\overleftarrow{\du}\cdot\overrightarrow{\du}\right)^k$ enables contraction between two rank-$k$ tensor-spinors, and has the properties:
$(\ast_k)u^\m=k\overleftarrow{\du}^\m(\ast_{k-1})$ and $(\ast_k)\overrightarrow{\du}^\m=(k+1)^{-1}u^\m(\ast_{k+1})$\,. The mass parameter:
\begin{equation}\label{tamm0}\m=\frac{1}{l}\left(n+\tf{D-4}{2}\right),\end{equation}
is uniquely fixed by gauge invariance~\cite{Metsaev:2006zy,Metsaev:2013wza}, where $l$ is the AdS radius.
The gauge symmetry of the Lagrangian~(\ref{f00}) is w.r.t.~a symmetric $\gamma$-traceless rank-$(n-1)$ tensor-spinor parameter:
\begin{equation}\label{tamm0.5}
\ve=\tf{1}{(n-1)!}\,\ve_{\m_1\ldots\m_{n-1}}u^{\m_1}\ldots u^{\m_{n-1}}, \qquad \dus\ve=0,
\end{equation}
while the triple $\g$-tracelessness condition~(\ref{tg1}) on the field translates in the operator formalism to:
\begin{equation}\label{tamm1} \dus\tr\ps=\tr\dus\ps=0.\end{equation}
Explicitly, the gauge transformations are given by:
\begin{equation}\label{tamm2}\d\ps=\grad\ve-\frac{1}{2l}\us\,\ve.\end{equation}
This can be verified by using the commutator~(\ref{commutator}), which reduces in AdS space to:
\begin{equation}\label{commutator-AdS}[\nb_\m,\nb_\n]=-\frac{1}{l^2}\left(u_{[\m} d_{\n]}+\tf{1}{2}\g_{\m\n}\right),\end{equation}
and the various commutators of the operators in $\mathbb{G}$ given the properties~(\ref{tamm0.5}) and~(\ref{tamm1}).

The metric-like description of higher-spin gauge fermions in flat-space is easily obtained by taking the limit $l\rightarrow\infty$ of the
gauge invariant system~(\ref{f00})--(\ref{commutator-AdS}). The degrees of freedom count in flat~\cite{Rahman:2015pzl} and AdS~\cite{Campoleoni:2017vds}
spaces are of course the same, and is given by:
\begin{equation}\label{dofF} \Delta_{\text{Metric}}=\binom{D+n-4}{n}f_D,\end{equation}
where $f_D$ for a Majorana fermion is given in Eq.~(\ref{fD-defined}), while for a Dirac fermion the value is twice as much.
Note that Eq.~(\ref{dofF}) counts the number of physical dynamical fields plus their conjugate momenta.
In AdS space, one of course gets the same number since the counting of dynamical equations, constraints and gauge freedom works in the same way.

As already mentioned in the Introduction, the $\g$-trace constraints~(\ref{tamm0.5})--(\ref{tamm1}) on the gauge parameter and the higher-spin fermionic field can be avoided by
recourse to other formulations. These include the non-local formulation~\cite{Francia:2002aa}, the BRST
formulation~\cite{Buchbinder:2004gp,Buchbinder:2007vq}, the higher-derivative compensator formulation~\cite{Francia:2007qt}, the quartet formulation~\cite{Buchbinder:2007ak}
and the non-minimal formulation with no higher derivatives~\cite{Campoleoni:2009gs}.

\end{appendix}

\bibliographystyle{ws-rv-van}

\begin{thebibliography}{99}

\bibitem{V-flat}
  M.~A.~Vasiliev,
  Yad.\ Fiz.\  {\bf 32} (1980) 855
   [Sov.\ J.\ Nucl.\ Phys.\  {\bf 32} (1980) 439].

\bibitem{AD}
  C.~Aragone and S.~Deser,
  Nucl.\ Phys.\ B {\bf 170} (1980) 329.

\bibitem{hygra}
  C.~Aragone and S.~Deser,
  Class.\ Quant.\ Grav.\  {\bf 1} (1984) L9.

\bibitem{Vasiliev:1986td}
  M.~A.~Vasiliev,
  Fortsch.\ Phys.\  {\bf 35} (1987) 741
   [Yad.\ Fiz.\  {\bf 45} (1987) 1784].

\bibitem{Lopatin:1987hz}
  V.~E.~Lopatin and M.~A.~Vasiliev,
  Mod.\ Phys.\ Lett.\ A {\bf 3} (1988) 257.

\bibitem{V-AdS}
  M.~A.~Vasiliev,
  Nucl.\ Phys.\ B {\bf 301} (1988) 26.

\bibitem{Vasiliev:2001wa}
  M.~A.~Vasiliev,
  Nucl.\ Phys.\ B {\bf 616} (2001) 106
   Erratum: [Nucl.\ Phys.\ B {\bf 652} (2003) 407]
  [hep-th/0106200].

\bibitem{deWit}
  B.~de Wit and D.~Z.~Freedman,
  Phys.\ Rev.\ D {\bf 21} (1980) 358.

\bibitem{Fronsdal:1978rb}
  C.~Fronsdal,
  Phys.\ Rev.\ D {\bf 18} (1978) 3624.

\bibitem{Fronsdal:1978vb}
  C.~Fronsdal,
  Phys.\ Rev.\ D {\bf 20} (1979) 848.

\bibitem{FF}
  J.~Fang and C.~Fronsdal,
  Phys.\ Rev.\ D {\bf 18} (1978) 3630.

\bibitem{Fang:1979hq}
  J.~Fang and C.~Fronsdal,
  Phys.\ Rev.\ D {\bf 22} (1980) 1361.

\bibitem{Rahman:2015pzl}
  R.~Rahman and M.~Taronna,
  ``From Higher Spins to Strings: A Primer,''
  arXiv:1512.07932 [hep-th].

\bibitem{Campoleoni:2017vds}
  A.~Campoleoni, M.~Henneaux, S.~Hörtner and A.~Leonard,
  JHEP {\bf 1702} (2017) 058
  [arXiv:1701.05526 [hep-th]].

\bibitem{Francia:2002aa}
  D.~Francia and A.~Sagnotti,
  Phys.\ Lett.\ B {\bf 543} (2002) 303
  [hep-th/0207002].

\bibitem{Bekaert:2003az}
  X.~Bekaert and N.~Boulanger,
  Phys.\ Lett.\ B {\bf 561} (2003) 183
  [hep-th/0301243].
  
\bibitem{Buchbinder:2004gp}
  I.~L.~Buchbinder, V.~A.~Krykhtin and A.~Pashnev,
  Nucl.\ Phys.\ B {\bf 711} (2005) 367.
  [hep-th/0410215].

\bibitem{Francia:2005bu}
  D.~Francia and A.~Sagnotti,
  Phys.\ Lett.\ B {\bf 624} (2005) 93.
  [hep-th/0507144].

\bibitem{Bekaert:2006ix}
  X.~Bekaert and N.~Boulanger,
  Commun.\ Math.\ Phys.\  {\bf 271} (2007) 723
  [hep-th/0606198].

\bibitem{Francia:2007qt}
  D.~Francia, J.~Mourad and A.~Sagnotti,
  Nucl.\ Phys.\ B {\bf 773} (2007) 203.
  [hep-th/0701163].
  
\bibitem{Buchbinder:2007ak}
  I.~L.~Buchbinder, A.~V.~Galajinsky and V.~A.~Krykhtin,
  Nucl.\ Phys.\ B {\bf 779} (2007) 155
  [hep-th/0702161].

\bibitem{Buchbinder:2007vq}
  I.~L.~Buchbinder, V.~A.~Krykhtin and A.~A.~Reshetnyak,
  Nucl.\ Phys.\ B {\bf 787} (2007) 211
  [hep-th/0703049].

\bibitem{Francia:2007ee}
  D.~Francia,
  Nucl.\ Phys.\ B {\bf 796} (2008) 77
  [arXiv:0710.5378 [hep-th]].

\bibitem{Campoleoni:2009gs}
  A.~Campoleoni, D.~Francia, J.~Mourad and A.~Sagnotti,
  Nucl.\ Phys.\ B {\bf 828} (2010) 405
  [arXiv:0904.4447 [hep-th]].

\bibitem{Campoleoni:2012hp}
  A.~Campoleoni, S.~Fredenhagen, S.~Pfenninger and S.~Theisen,
  J.\ Phys.\ A {\bf 46} (2013) 214017
  [arXiv:1208.1851 [hep-th]].

\bibitem{Fredenhagen:2014oua}
  S.~Fredenhagen and P.~Kessel,
  J.\ Phys.\ A {\bf 48} (2015) no.3,  035402
  [arXiv:1408.2712 [hep-th]].

\bibitem{Campoleoni:2014tfa}
  A.~Campoleoni and M.~Henneaux,
  JHEP {\bf 1503} (2015) 143
  [arXiv:1412.6774 [hep-th]].

\bibitem{Boulanger:2015ova}
  N.~Boulanger, P.~Kessel, E.~D.~Skvortsov and M.~Taronna,
  J.\ Phys.\ A {\bf 49} (2016) no.9,  095402
  [arXiv:1508.04139 [hep-th]].

\bibitem{Konstein:1989ij}
  S.~E.~Konstein and M.~A.~Vasiliev,
  Nucl.\ Phys.\ B {\bf 331} (1990) 475.

\bibitem{Vasiliev:1990vu}
  M.~A.~Vasiliev,
  Class.\ Quant.\ Grav.\  {\bf 8} (1991) 1387.

\bibitem{Vasiliev:1992av}
  M.~A.~Vasiliev,
  Phys.\ Lett.\ B {\bf 285} (1992) 225.

\bibitem{Sezgin:1998gg}
  E.~Sezgin and P.~Sundell,
  JHEP {\bf 9811} (1998) 016
  [hep-th/9805125].

\bibitem{Sezgin:1998eh}
  E.~Sezgin and P.~Sundell,
  In *Goeteborg 1998, Novelties in string theory* 241-269
  [hep-th/9903020].

\bibitem{Sezgin:2002ru}
  E.~Sezgin and P.~Sundell,
  JHEP {\bf 0207} (2002) 055
  [hep-th/0205132].

\bibitem{Engquist:2002vr}
  J.~Engquist, E.~Sezgin and P.~Sundell,
  Class.\ Quant.\ Grav.\  {\bf 19} (2002) 6175
  [hep-th/0207101].

\bibitem{Sezgin:2012ag}
  E.~Sezgin and P.~Sundell,
  J.\ Phys.\ A {\bf 46} (2013) 214022
  [arXiv:1208.6019 [hep-th]].

\bibitem{Alkalaev:2001mx}
  K.~B.~Alkalaev,
  Phys.\ Lett.\ B {\bf 519} (2001) 121
  [hep-th/0107040].

\bibitem{Alkalaev:2006hq}
  K.~B.~Alkalaev,
  Theor.\ Math.\ Phys.\  {\bf 149} (2006) 1338
   [Teor.\ Mat.\ Fiz.\  {\bf 149} (2006) 47]
  [hep-th/0501105].

\bibitem{Sorokin:2008tf}
  D.~P.~Sorokin and M.~A.~Vasiliev,
  Nucl.\ Phys.\ B {\bf 809} (2009) 110
  [arXiv:0807.0206 [hep-th]].

\bibitem{Z}
  Y.~M.~Zinoviev,
  Nucl.\ Phys.\ B {\bf 808} (2009) 185
  [arXiv:0808.1778 [hep-th]].

\bibitem{Z2}
  Y.~M.~Zinoviev,
  Nucl.\ Phys.\ B {\bf 821} (2009) 21
  [arXiv:0904.0549 [hep-th]].

\bibitem{Skvortsov:2010nh}
  E.~D.~Skvortsov and Y.~M.~Zinoviev,
  Nucl.\ Phys.\ B {\bf 843} (2011) 559
  [arXiv:1007.4944 [hep-th]].

\bibitem{Hallowell:2005np}
  K.~Hallowell and A.~Waldron,
  Nucl.\ Phys.\ B {\bf 724} (2005) 453
  [hep-th/0505255].

\bibitem{Metsaev:2006zy}
  R.~R.~Metsaev,
  Phys.\ Lett.\ B {\bf 643} (2006) 205
  [hep-th/0609029].

\bibitem{Metsaev:2013wza}
  R.~R.~Metsaev,
  arXiv:1311.7350 [hep-th].

\bibitem{Metsaev:2003cu}
  R.~R.~Metsaev,
  Phys.\ Lett.\ B {\bf 590} (2004) 95
  [hep-th/0312297].

\bibitem{Freedman:2012zz}
  D.~Z.~Freedman and A.~Van Proeyen,
  ``Supergravity,''
  Cambridge Univ.~Press, 2012.
  ISBN-10: 0521194016

\bibitem{Troncoso}
  O.~Fuentealba, J.~Matulich and R.~Troncoso,
  JHEP {\bf 1509} (2015) 003
  [arXiv:1505.06173 [hep-th]].

\bibitem{ours}
  M.~Henneaux, G.~Lucena G\'omez and R.~Rahman,
  ``The uniqueness of hypergravity,''
  to appear.

\bibitem{Henneaux:2012wg}
  M.~Henneaux, G.~Lucena G\'omez and R.~Rahman,
  JHEP {\bf 1208} (2012) 093
  [arXiv:1206.1048 [hep-th]].

\bibitem{Henneaux:2013gba}
  M.~Henneaux, G.~Lucena G\'omez and R.~Rahman,
  JHEP {\bf 1401} (2014) 087
  [arXiv:1310.5152 [hep-th]].

\bibitem{Fradkin:1986ka}
  E.~S.~Fradkin and M.~A.~Vasiliev,
  Annals Phys.\  {\bf 177} (1987) 63.

\bibitem{Fradkin:1987ks}
  E.~S.~Fradkin and M.~A.~Vasiliev,
  Phys.\ Lett.\ B {\bf 189} (1987) 89.

\bibitem{Fradkin:1986qy}
  E.~S.~Fradkin and M.~A.~Vasiliev,
  Nucl.\ Phys.\ B {\bf 291} (1987) 141.

\bibitem{Curtright:1979uz}
  T.~Curtright,
  Phys.\ Lett.\  {\bf 85B} (1979) 219.


\end{thebibliography}

\end{document}